\documentclass[a4paper,preprintnumbers,floatfix,superscriptaddress,prx,showpacs,notitlepage,longbibliography]{revtex4-2}
\usepackage[english]{babel}
\usepackage{ucs}
\usepackage[utf8x]{inputenc}
\usepackage{graphicx}
\usepackage{dcolumn}
\usepackage{bm}
\usepackage{bbm}
\usepackage{amsmath}
\usepackage{amssymb}
\usepackage{mathrsfs}
\usepackage{amsfonts}
\usepackage[table, dvipsnames]{xcolor}
\usepackage[colorlinks=true,citecolor=PineGreen,linkcolor=PineGreen,urlcolor=black]{hyperref}
\usepackage[normalem]{ulem}

\newcommand{\ket}[1]{|#1\rangle}

\newcommand{\ex}[1]{\mathrm{e}^{#1}}
\newcommand{\Ket}[1]{\left|#1\right>}
\newcommand{\Bra}[1]{\left<#1\right|}
\newcommand{\BraKet}[2]{\left<#1|#2\right>}
\newcommand{\KetBra}[2]{|#1\rangle\langle#2|}
\newcommand{\trace}[1]{\mathrm{Tr}\left(#1\right)}

\graphicspath{{./Figs/}}

\begin{document} 

\title{Generating nonlinearities from conditional linear operations, squeezing and measurement for quantum computation and super-Heisenberg sensing }

\author{Mattias T. Johnsson} 
\affiliation{Centre for Engineered Quantum Systems, Department of Physics and Astronomy,
Macquarie University, Sydney, New South Wales 2109, Australia }
\author{Pablo M. Poggi} 
\affiliation{Center for Quantum Information and Control, University of New Mexico, MSC07-4220, Albuquerque, New Mexico 87131-0001, USA}
\author{Marco A. Rodriguez}
\affiliation{Instituto de Investigaciones en Matematicas Aplicadas y Sistemas, Universidad Nacional Auntonoma de Mexico, CDMX. 01000, Mexico}
\author{Rafael N. Alexander} 
\affiliation{Center for Quantum Information and Control, University of New Mexico, MSC07-4220, Albuquerque, New Mexico 87131-0001, USA}
\affiliation{Centre for Quantum Computation and Communication Technology,School of Science, RMIT University, Melbourne, Victoria 3001, Australia}
\author{Jason Twamley} 
\affiliation{Quantum Machines Unit, Okinawa Institute of Science and Technology Graduate University, Okinawa 904-0495, Japan}
\affiliation{Centre for Engineered Quantum Systems, Department of Physics and Astronomy, Macquarie University, Sydney, New South Wales 2109, Australia }

\date{\today}

\begin{abstract}
Large optical nonlinearities can have numerous applications, ranging from the generation of cat-states for optical quantum computation, through to quantum sensing where the sensitivity exceeds Heisenberg scaling in the resources. However, the generation of ultra-large optical nonlinearities has proved immensely challenging experimentally. We describe a novel protocol where one  can effectively generate large optical nonlinearities via  the conditional application of a { linear operation} on an optical mode by an ancilla mode, followed by a measurement of the ancilla and corrective operation on the probe mode. Our protocol can generate high quality optical Schr{\"{o}}dinger cat states useful for optical quantum computing and can be used to perform sensing of an unknown rotation {or} displacement in phase space, with super-Heisenberg scaling in the resources. We finally describe a potential experimental implementation using atomic ensembles interacting with optical modes via the Faraday effect.
\end{abstract}

\maketitle

\section{Introduction} 
Optical nonlinearities, and in particular the Kerr nonlinear oscillator, have been the focus of much research within quantum optics since the first seminal investigations by Milburn and Holmes \cite{MilburnandHolmes1986} and Yurke and Stoler \cite{Yurke1986}. Nonlinear quantum oscillators and the highly non-classical cat states they can produce have found numerous applications including studying the fundamentals of decoherence \cite{Zurek2003}, improved schemes for metrology \cite{Luis1992BreakingMeasurements, Munro2002, Luis2004NonlinearLimit, Beltran2005BreakingDetectors,Boixo2007GeneralizedEstimation, Boixo2008Quantum-limitedStates, Woolley2008NonlinearResonators, Boixo2008QuantumEntanglement, Gross2010NonlinearLimit, Rivas2010PrecisionSchemes, Napolitano2010NonlinearInterface, Napolitano2011Interaction-basedLimit, Hall2012DoesTheory, Joo2012QuantumStates, Berrada2013QuantumModes, Berrada2013QuantumShifts, Cheng2014QuantumShifts, Luis2015NonlinearMetrology, Luis2016NonlinearMetrology, Beau2017NonlinearSystems, Wei2017ImprovingShifter, Florez2018TheInterferometer, Nie2018ExperimentalState, Tsarev2019NonlinearSolitons, Zhang2019NonlinearStates}, as well as for quantum computation \cite{Munro2005WeakComputation, Albert2019Pair-catNonlinearity,Grimsmo2020QuantumCodes}. In particular, research has shown that nonlinear quantum systems can provide a metrological precision that scales better than the so-called Heisenberg scaling in the estimation of a parameter $\phi$. Standard quantum limit (SQL) and Heisenberg quantum limit (HL) metrology  schemes result in an imprecision $\delta\phi$ that scales with the resource $\bar{n}$ as $\delta \phi\sim 1/\sqrt{\bar{n}}$ or $1/\bar{n}$, respectively.  

Although nonlinear  Kerr-type oscillators have been intensively studied theoretically, experimentally implementing them has proved extremely challenging. The degree of nonlinearity that can be engineered in most atomic or optical systems is too small, or is associated with too much loss, to be useful. Recently, superconducting quantum devices have proved capable of generating Kerr-type quantum states in the microwave domain \cite{Kirchmair2013ObservationEffect}, but their generation in the optical domain remains problematic.

In this work we show that, curiously, one can {\em imprint} a nonlinear Hamiltonian (in our case a Kerr) on an optical mode (which we will denote as the Probe mode) using only a conditional {\bf linear} operation from an ancilla mode which is then measured. This measurement implements a Kraus operation on the primary mode which comprises of both unitary and non-unitary components and in a suitable limit we find that the Kraus operation is almost of a pure Kerr type. We show how tailoring this Kraus operation  
\begin{enumerate}
    \item can be used to perform super-Heisenberg  sensing of an unknown rotation in phase space with an imprecision which scales as $\delta \theta\sim 1/\bar{n}^{3/2}$,
    \item can be used to perform super-Heisenberg sensing of an unknown displacement in phase space with $\delta x\sim 1/\bar{n}^{3/2}$,
    \item can be used to engineer near perfect non-classical Kerr cats for use in optical quantum computation and metrology, and
    \item can be implemented using optical modes interacting with atomic ensembles via the Faraday effect.
\end{enumerate}

In Section \ref{basicidea} we describe the fundamental idea behind the scheme, which is graphically depicted in Fig.~\ref{fig:abstractcircuit1}. In Section \ref{enhancedmetrology}, we introduce the reader to some principals of quantum metrology and in \ref{quick}, we give a quick introduction to quantum Fisher information. In Section \ref{QFIensemble}, we extend the quantum Fisher information to post-measurement ensembles and in Sections \ref{superh}, and \ref{bootstrap}, we describe how to perform super-Heisenberg metrology. In Section \ref{preparation} we describe how to use our protocol to generate Schr{\"{o}}dinger cat and compass states with high fidelity and finally, in Section \ref{sect:atomic_prot} we describe a potential experimental implementation of our protocol. 

%
%
%

\section{Outline of our protocol to generate nonlinear dynamics} 
\label{basicidea}
The protocols we develop in this paper are primarily based on the innocuous Gaussian integral 
\begin{equation}
    \int_{-\infty}^{+\infty}\,dy\, e^{-a y^2+b y}=\sqrt{\frac{\pi}{a}}\, e^{b^2/(4a)},
    \label{gaussianintegral}
\end{equation}
which converges provided ${\rm Re}(a)>0$.
This integration has the interesting property that the parameter $b$, which appears linearly in the exponential of the integrand on the left, ends up appearing quadratically in the exponential on the right. Unitary operators can be written as the exponential of an Hermitian generator. We will make use of the curious property of Eq.~(\ref{gaussianintegral}) to essentially {\em square} the generator. In particular we will show how it is possible to {\em bootstrap} the typical harmonic oscillator generator $\hat{H}_{ho}\sim \hat{a}^\dagger \hat{a}\sim \hat{n}$ to become that of the nonlinear Kerr oscillator $\hat{H}_{ko}\sim\hat{n}^2$. We will see that this bootstrapping can only be achieved  approximately with realistic resources, but interestingly it can be achieved deterministically even though a measurement is involved.

\onecolumngrid

\begin{figure}
\centering
\setlength{\unitlength}{1cm}
\begin{picture}(8,4)
\put(-3,0){ \includegraphics[width=.75 \linewidth]{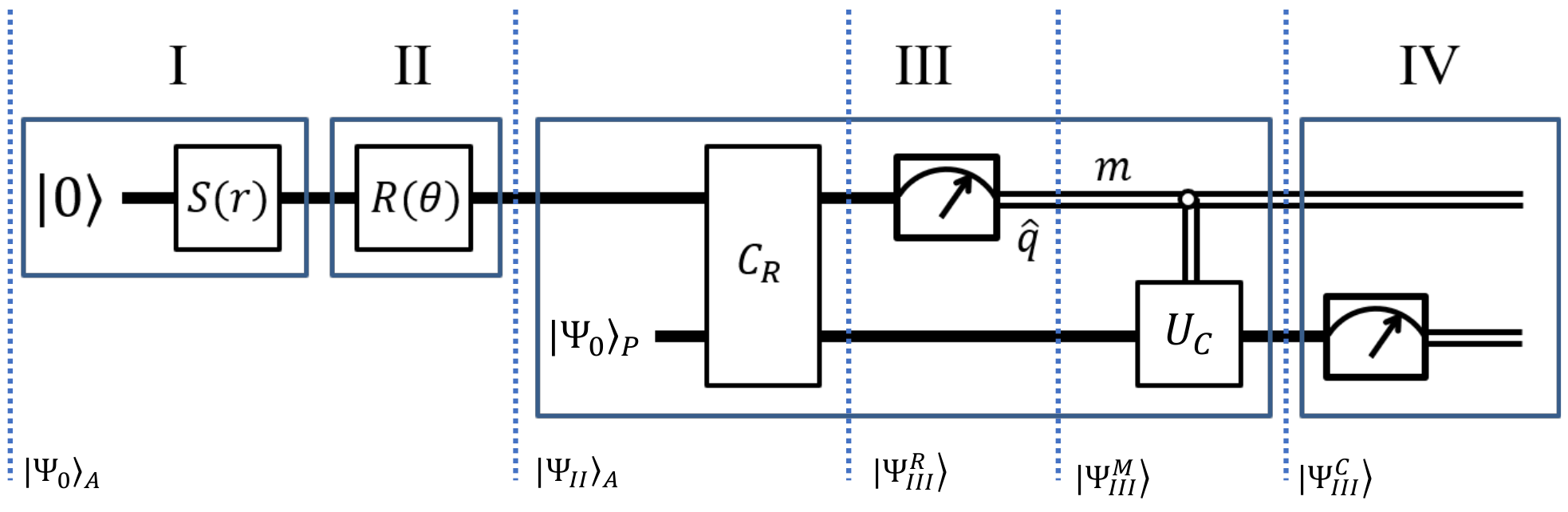}}
\end{picture}
\caption{Basic protocol: Both the top and bottom lines of the circuit represent bosonic modes which we label as the Ancilla (top), and Probe (bottom) modes. In Stage I we initialize the Ancilla mode in a squeezed vacuum state, where $S(r)$ is the squeezing operator. In Stage II we implement a rotation of the Ancilla mode by $R(\theta)$. In Stage III, we implement ``Kerr teleportation'' onto the input state $\ket{\psi}$ (which we assume is a coherent state), via the cross-rotation gate  $C_R$, (which rotates the Probe mode by an amount dependent on the momentum of the Ancilla mode, e.g. generated by $\hat{H}_{R}=\tilde{g}\, \hat{p}_A\otimes \hat{n}_P$), and a position quadrature measurement of the first mode. This has the upshot of implementing both linear $\sim \hat{n}_P$, and quadratic $\sim \chi\, \hat{n}_P^2$, operations on the Probe mode, where the Kerr strength $\chi$ is function of $r, \theta$ and $g$.  We allow for a measurement outcome dependent unitary $U_C$ at the conclusion of this teleportation procedure to {\em undo} the linear rotations leaving only Kerr-like evolution with some decay. Finally, the Probe mode can be measured in Stage IV. Quantum states referred to in the text at various points in the circuit are labelled by dotted vertical lines. }
    \label{fig:abstractcircuit1}
\end{figure}

We introduce this protocol via Fig.~\ref{fig:abstractcircuit1}, which is the basic description of the {\em bootstrap} protocol. We begin with two bosonic modes as shown in Fig.~\ref{fig:abstractcircuit1}, with the top (bottom) mode denoted as the Ancilla (Probe) modes. We introduce the following single and two mode operators
\begin{eqnarray}
\hat{S}(r)&=&e^{-r\,(\hat{a}_A^{2} - \hat{a}_A^{\dagger 2})/2},\\ \label{eq:sqgate}
\hat{R}(\theta)&=&e^{i\,\theta \hat{n}_A},\\\label{eq:rotgate}
\hat{Sh}(\beta)&=&e^{-i\,\beta\, \hat{p}^2_A },\\\label{eq:sheargate}
\hat{C}_R &=& e^{-i\,g\, \hat{p}_A \otimes\, \hat{n}_{P} },\label{eq:crotgate}
\end{eqnarray}
where the $A(P)$, subscript indicates operators acting on the Ancilla (Probe) modes. These operators we name as the $S(r)$: squeezing, $R(\theta)$: rotation, $Sh(\beta)$: shear, and $C_R$: cross-rotation operators. Referring to Fig. \ref{fig:abstractcircuit1}, we begin by considering Stage I and II involving the Ancilla mode where, for simplicity, initially we will apply $\hat{Sh}(\beta)$ at Stage II rather than $\hat{R}(\theta)$. We will generalise to the case including rotation later. We can write the state of the Ancilla mode after Stage II as
\begin{eqnarray}
   \Ket{\psi_{II}}_A\equiv \hat{Sh}(\beta)\,\hat{S}(r)\,&&|0\rangle_A\\
    &&=e^{-i \beta \hat{p}_A^2}\hat{S}(r)\Ket{0}_A\\&&= \mathcal{N}_{II} \int dp\: e^{-\frac{p^2}{2\sigma^2}} e^{-i\beta p^2} \Ket{p}_A\label{eqn132}
\end{eqnarray}
where $\Ket{p}_A$ is the eigenstate of $\hat{p}_A$, the momentum operator of the Ancilla mode,  $\sigma^2=e^{2 r}$, and $\mathcal{N}_{II}$ is a normalization constant. We next bring in the initial state of the Probe mode $\Ket{\psi}_P\equiv \Ket{\alpha}_P$, taking it to be in a coherent state of magnitude $\alpha$, and apply the cross-rotation gate to obtain
\begin{eqnarray}
    \Ket{\Psi_{III}^R}&\equiv& e^{-i g\hat{p}_A\otimes \hat{n}_P}\Ket{\psi_{II}}_A \otimes \Ket{\alpha}_P\\
    &=&\mathcal{N}_{II}\int dp\: e^{-p^2\left(\frac{1}{2\sigma^2} + i\beta\right)} e^{-i\,g\,p\,\hat{n}_P} \Ket{p}_A\otimes \Ket{\alpha}_P.\label{IIIR}
\end{eqnarray}
Next, in Stage III we apply a position measurement on the Ancilla mode, and if the result of that measurement is $m$, and given $\BraKet{m}{p}_A=e^{imp} { /\sqrt{2\pi}}$,  we obtain the post-measured state of the Probe to be
\begin{equation}
    \Ket{\Psi_{III}^{M}} = \mathcal{N}_{IIIM} \int dp\, e^{-p^2\left(\frac{1}{2\sigma^2} + i\beta\right)} e^{ip\left(m-g \hat{n}_P\right)} \Ket{\alpha}_P,\label{IIIM}
\end{equation}
where ${\cal{N}}_{IIIM} = 1/(\pi^{1/4} \sqrt{2\pi \sigma})$.
We can now proceed to integrate over the integration variable $p$. For this, we use the known integral
\begin{equation}
    \int\limits_{-\infty}^{+\infty} dp\, e^{-ap^2+bp}=\sqrt{\frac{\pi}{a}}\exp{\left(\frac{b^2}{4a}\right)}\label{doesthishaveaname}
\end{equation}
\noindent which holds only if $\mathrm{Re}(a)>0$. Identifying $a=1/2\sigma^2+i\beta$ and $b=i(m-g\,\hat{n}_P)$, we obtain for the post-measurement state of the probe mode
\begin{eqnarray}
\Ket{\Psi_{III}^M} &=& \mathcal{N}_{IIIM}^\prime \exp{\left(-\frac{(g \hat{n} -m)^2}{4\left(\frac{1}{2\sigma^2}+i\beta\right)}\right)}\Ket{\alpha}\label{ec:output1}\\
&=&\mathcal{N} \hat{U}(\beta)\hat{U}_c(\beta,m) \hat{K}(\beta)\hat{K}_c(\beta,m)\Ket{\alpha}
\label{ec:output2}
\end{eqnarray}
where we have now dropped the subscript $P$ referring to the Probe mode. In the final expression (\ref{ec:output2}) we have decomposed the propagator into a product of unitary and non-unitary operations, where:
\begin{eqnarray}
\hat{U}(\beta)&=&\exp{\left(i\, \frac{\beta g^2}{\mu}\;\hat{n}^2\right)} \label{eq:Ubeta} \\
\hat{K}(\beta)&=& \exp{\left(-\frac{g^2}{2\sigma^{2}\mu}\; \hat{n}^2\right)} \label{eq:Kbeta} \\
\hat{U}_c(\beta,m)&=&\exp{\left(i \frac{2\beta m g }{\mu}\;\hat{n}\right)} \label{Ucequation} \label{eq:Ucbetam} \\
\hat{K}_c(\beta,m)&=& \exp{\left( -\frac{2\,m\,g}{2\sigma^{2}\mu}\; \hat{n} \right)}, \label{eq:Kcbetam} 
\end{eqnarray}
where $\mu=4(\beta^2+1/(4\sigma^4))$ and $\cal N$ is a normalisation factor. We notice that the conditioned evolution of the Probe state involves  unitary and non-unitary operations $\hat{U}(\beta)$, and $\hat{K}(\beta)$, which do {\em not} depend on the measurement result $m$, while the remaining two $\hat{U}_c(\beta,m)$ and $\hat{K}_c(\beta,m)$, {\em do} depend on $m$. 

We now reach our first important observation: that the unitary $\hat{U}(\beta)$, is a deterministic pure Kerr evolution where the Kerr strength $\chi\equiv \beta g^2/\mu$ now depends on $\beta$ and $\sigma$, the shear and squeezing parameters of the operations on the  Ancilla mode. We also see that all the operations (\ref{eq:Ubeta})--(\ref{eq:Kcbetam}) commute irrespective of the parameter values. In the limit of infinite squeezing $\sigma\rightarrow \infty$, we have $\mu\rightarrow 4\beta^2$, and the two non-unitary operators collapse to the identity while $\chi\rightarrow g^2/4\beta$, and $\hat{U}_c$ is an $m$-dependent rotation with phase $\phi=mg/2\beta$. In this limit we are left with pure unitary evolution that consists of a deterministic non-linear rotation and a measurement-dependent linear rotation. In the final right-most section of III in Fig.~\ref{fig:abstractcircuit1} we assume we can apply a {\em correction} unitary on the Probe mode which depends on the value of the measurement outcome $m$. 

In the above simplified initial description of the protocol we applied the shearing operator $\hat{Sh}(\beta)$, rather than a rotation operator $\hat{R}(\theta)$. Use of the shearing operation permits a relatively straightforward illustration of the main principals of how the protocol operates and the resulting analytic expressions are compact.  However, it is more physically relevant to use the rotation operator $\hat{R}$. We note that the combination of squeeze and shear operations can be decomposed as a combination of squeeze and rotation operations and this alternative parameterisation is shown in Fig.~\ref{fig:abstractcircuit1}. We will make use of this latter description in the remainder of the work below. To find this alternative parameterisation we use the Siegel upper half space representation of Gaussian pure states, where the state is be represented by a complex number $z= v + i u$, where $u >0$~\cite{Menicucci2011GraphicalStates}. We note that the symplectic matrices corresponding to squeezing and rotation can be written as
\begin{align}
\begin{pmatrix} 0 & -1 \\ 1 & 0 \end{pmatrix}\begin{pmatrix} \cos \theta & -\sin \theta \\ \sin \theta & \cos \theta \end{pmatrix} \begin{pmatrix} e^{-r} & 0 \\ 0 & e^{r} \end{pmatrix} = \begin{pmatrix} a & b \\ c & d \end{pmatrix}
\end{align}
where the leftmost matrix implements an extra $\pi/2$, rotation which is responsible for interchanging the roles of the position and momentum bases, and the state gets squeezed in position for $r>0$. 
Then
\begin{equation}
     \frac{c + i d}{a + i b} = z = v + i u,\label{holomorpheqn}
\end{equation}
 where $u=1/\Xi$, $v=-\sin{2\theta} \sinh{2r}/\Xi$, and $\Xi=e^{-2r} \sin^{2} \theta + e^{2r} \cos^{2}\theta$.

One can express the state output from stage II as
\begin{align}
    \ket{\psi_{II}}_A &= \hat{R}(\theta) \hat{S}(r) \ket{0} \\
    &= \left( \frac{u}{\pi}\right)^{1/4} \int \mathrm{d}p\;  e^{-\frac{1}{2} p^{2}(u-iv)}\; \ket{p}_{A},
\end{align}
and repeating the steps from (\ref{eqn132}) -- (\ref{ec:output2}), we can show 
\begin{align}
\frac{1}{2\sigma^{2}} = u, \label{eq:sigma_beta_u_v1}\\
\beta = -v,
\label{eq:sigma_beta_u_v2}
\end{align}
and using (\ref{eq:sigma_beta_u_v1}) and (\ref{eq:sigma_beta_u_v2}) one can re-express the parameters  $(\sigma,\beta)\rightarrow (r,\theta)$. We will primarily use the $(r,\theta)$ parameterisation in the remainder of the work below.
\vspace{1cm}

In the remainder of the paper we will explore two main variations of this circuit: 
\begin{description}
\item[Enhanced Quantum Metrology] If  we assume we have imperfect knowledge of one parameter in the circuit, e.g. of the angle $\theta$, and wish to estimate the value of $\theta$, we first describe how, using the circuit in Fig.~\ref{fig:abstractcircuit1}, we can perform this estimate with a precision that scales as $\Delta\theta\sim 1/\bar{n}_{P}^{3/2}$, where $\bar{n}_P$ is the mean photon number of the input Probe coherent state. This scaling in precision is faster than the typical Heisenberg scaling for estimating $\theta$, which normally scales as $\Delta\theta\sim 1/\bar{n}_{P}$. We describe how the circuit shown in Fig.~\ref{fig:abstractcircuit1} can be used to estimate $\theta$, the parameter of a phase rotation of a mode, and alternatively, how to estimate $\kappa$, which parameterises displacements of a mode, each with a precision that scales as $\sim1/\bar{n}_P^{3/2}$. We note that in this latter example, which is often used for force sensing, the standard Heisenberg limit scales as $\Delta\kappa\sim 1/\sqrt{\bar{n}}$. Thus our improvement in displacement metrology over the normal Heisenberg limit is substantial \cite{Dalvit2006}. We discuss this scaling further in section \ref{bootstrap}.
\item [Non-Gaussian State Preparation]
Next we will assume we have full information of all the parameters in the circuit. 
With full knowledge of these parameters we are able to apply complete nonlinear correction so that in the high-squeezing limit we are left with a deterministic pure Kerr evolution $\hat{U}(\theta)=\exp(i\, (g^2/4)\cot(\theta)\; \hat{n}^2)$, which, curiously, has a Kerr strength which is a highly nonlinear function of $\theta$. We can use this to produce non-classical states of the Probe mode and, in particular, with infinite squeezing and pure-Kerr evolution, we can target the generation of a Yurke-Stoler cat state \cite{Yurke1986}. A more realistic scenario, using finite squeezing,  will result in imperfect preparation of such non-Gaussian quantum states. In Section \ref{preparation}, we study the preparation fidelities that can be achieved using this scheme.
\end{description}

\section{Performing Enhanced Metrology} \label{enhancedmetrology}
We now develop the quantum circuit outlined in Fig.~\ref{fig:abstractcircuit1} to perform parameter estimation as outlined at the end of Section \ref{basicidea}. We assume that we have complete knowledge of the parameters $(r, g, m, \alpha_P)$ denoting the Ancilla mode squeezing, strength of the cross-rotation gate $C_R$, measurement outcome $m$, and the  parameter $\alpha_P$ describing the coherent-state input to the Probe mode, respectively. We assume that the parameter $\theta$ is set to a particular known base value (which we denote as $\theta$), and we are interested in estimating changes in $\theta\rightarrow \theta+\delta \theta$ about this base value  with precision. We also assume we can apply a {\em correction} rotation unitary at the end of Stage III of the circuit shown in Fig.~\ref{fig:abstractcircuit1},  or $\hat{U}_c(f(r,g,\theta),m)$ in Eq.  (\ref{eq:Ucbetam}), where $f$ is a deterministic function of these known parameters. This correction unitary aims to reduce the stochastic effects of the measurement on the Probe. One finally performs a measurement on the Probe mode in order to estimate the unknown parameter change $\delta\theta$. As a measure of the resources required to achieve a particular precision in estimation we will make use of the Quantum Fisher Information (QFI). The QFI is typically defined for unitary channels, where the effect on the final state by a change in the parameter is unitary. However, in the circuit shown in Fig.~\ref{fig:abstractcircuit1}, information about $\delta\theta$ is found not only in the final conditioned state of the Probe mode $\hat{\rho}_P^m$, but also in the classical measurement results $m$.  We have to then expand the normal unitary QFI to encompass the ensemble of joint classical/quantum outputs states $\left\{ m, \hat{\rho}_P^m\right\}$.

\vspace{1cm}

\subsection{Quick Review of Quantum Fisher Information} \label{quick}
In quantum metrology one aims to statistically estimate the value of a parameter in the system using an unbiased estimator. From the quantum Cram\'{e}r-Rao theorem the Quantum Fisher Information (QFI) provides a lower limit on the variance of such an estimator \cite{Fisher1925TheoryEstimation, Helstrom1967MinimumStatistics, Braunstein1994StatisticalStates,Braunstein1996GeneralizedInvariance, Giovannetti2011AdvancesMetrology}. A larger value of the QFI implies higher precision parameter estimation and one can study the dependence of the QFI on various quantum resources, e.g. the average photon number of the input Probe mode.  We now give the reader a quick overview of the properties of the QFI illustrated with a number of examples before addressing the QFI of a channel of pure states conditioned on classical measurement outcomes.

As mentioned above the precision of a statistical estimation of a parameter $\theta$ can be studied in terms of the (classical) Fisher information (FI), $F(\theta)$, which determines the Cram\'{e}r-Rao bound for the variance of an unbiased estimator

\begin{equation}
    \Delta \theta \geq \Delta \theta_{CR} = \frac{1}{\sqrt{\nu F(\theta)}},
\end{equation}
where $\nu$ quantifies the total number of repetitions of the estimation. 

The FI can be upper bounded by the Quantum Fisher Information $F_Q$ (QFI). The QFI is a function of a family of parameterised quantum states $\{\hat{\rho}(\theta)\}$,
\begin{equation}
    F_Q[\hat{\rho}(\theta)]=\trace{\hat{\rho}(\theta) \hat{L}^2}, \,\,\,\, \mathrm{where}\ \frac{\partial \hat{\rho}(\theta)}{\partial \theta} \equiv \frac{1}{2}\left(\hat{\rho}(\theta) \hat{L} + \hat{L} \hat{\rho}(\theta)\right),
    \label{ec:qfi_sld}
\end{equation}
and where $\hat{L}$ is a $\theta$-dependent Hermitian operator called the symmetric logarithmic derivative (SLD). When $\hat{\rho}(\theta)$ is pure, the SLD and the QFI are easy to calculate (see \cite{Pezze2018QuantumEnsembles, Paris2009QuantumTechnology}). Because $\rho_\theta=\rho_\theta^2$, we have
\begin{equation}
    \frac{\partial \hat{\rho}(\theta)}{\partial \theta}=\frac{\partial}{\partial \theta}\left(\hat{\rho}^2(\theta)\right)=\frac{\partial \hat{\rho}(\theta)}{\partial \theta}\hat{\rho}(\theta) + \hat{\rho}(\theta) \frac{\partial \hat{\rho}(\theta)}{\partial \theta},
\end{equation}

\noindent which immediately gives, from Eqn. (\ref{ec:qfi_sld}), 
\begin{equation}
    \hat{L}=2\frac{\partial \hat{\rho}(\theta)}{\partial \theta}=\KetBra{\partial_\theta \psi_\theta}{\psi_\theta}+\KetBra{\psi_\theta}{\partial_\theta \psi_\theta},
\end{equation}
\noindent where we have expressed $\hat{\rho}(\theta) = \KetBra{\psi_\theta}{\psi_\theta}$ and denoted $\frac{\partial}{\partial \theta}\rightarrow \partial_\theta$. We can thus write down the QFI for pure states,
\begin{equation}
    F_{Q}[\hat{\rho}(\theta)]=\Bra{\psi_\theta}\hat{L}^2\Ket{\psi_\theta}=4\left(\BraKet{\partial_\theta \psi_\theta}{\partial_\theta \psi_\theta}-\lvert \BraKet{\psi_\theta}{\partial_\theta \psi_\theta}\rvert^2\right).
    \label{ec:qfi_pure}
\end{equation}
If the parameterised pure states $\{\Ket{\psi_\theta}\}$ are generated by a $\theta$-dependent unitary transformation acting on a fiducial state $\Ket{\psi_0}$, i.e. if $\Ket{\psi_\theta}=\exp\left(-i\hat{G}\,\theta\right)\,\Ket{\psi_0}$, then the expression above reduces to
\begin{equation}
    F_Q[\rho_\theta]= F_Q[\Ket{\psi_0},\hat{G}]=4\left(\Bra{\psi_0}\,\hat{G}^2\Ket{\psi_0} - \Bra{\psi_0}\,\hat{G}\Ket{\psi_0}^2\right),
    \label{ec:qfi_hamilt}
\end{equation}
which is four times the variance of the generator $\hat{G}$ in that fiducial state $\Ket{\psi_0}$, which we will denote as $(\Delta \hat{G})\rvert^2_{\psi_0}$.

For illustrative purposes, we now consider some applications of these QFI relations with respect to metrology. We first consider estimating an \textit{unknown phase} imprinted on the state of a single quantum bosonic mode prepared in the fiducial coherent state $\Ket{\alpha}$, which is subject to an unknown {\em linear} phase shift via the operation $R(\theta)=\ex{-i\theta \hat{n}}$, where $\hat{n}=\hat{a}^\dagger \hat{a}$. From Eqn. (\ref{ec:qfi_hamilt}), we find the pure state QFI as 
$
    F_Q[\Ket{\alpha},\hat{n}]=4(\Delta \hat{n})\rvert^2_\alpha=4|\alpha|^2 = 4 \bar{n},
    \label{ec:qfi_sql}
$
using the notation for the variance introduced above. Here we have also introduced $\bar{n}$ as the mean number occupation of the fiducial state. We will be particularly focused  on analyzing the scaling of the QFI with $\bar{n}$ for different types of metrology protocols, treating $\bar{n}$ as a quantification of the quantum resource. In this case, the scaling $F_Q[\Ket{\alpha},\hat{n}]\sim \bar{n}$, represents the \textit{standard quantum limit} (SQL) for phase estimation. The SQL for phase estimation can be beaten by imprinting linear phase shifts on squeezed states, as shown, for example, in  \cite{Monras2006OptimalStates}. An input fiducial state which is a squeezed vacuum state $\Ket{\phi_r}=\hat{S}(r)\Ket{0}$, with mean photon number $\bar{n}$,  yields a QFI $
    F_Q[\Ket{\phi_r},\hat{n}]=4(\Delta \hat{n})\rvert^2_{\phi_r}=8\sinh(r)^2\cosh(r)^2
    \label{ec:qfi_hl}=8(\bar{n}^2 + \bar{n})$,
and thus leads to a better scaling of the QFI with $\bar{n}$, or the so-called {\em Heisenberg} scaling of the estimation of the phase where $\Delta\theta\sim 1/\bar{n}$.

Finally, we look at the case of generating a \textit{nonlinear} phase shift on a coherent state, that is applying  a transformation like $\ex{-i\theta \hat{n}^2}$ to $\Ket{\alpha}$. Following Ref. \cite{Rivas2010PrecisionSchemes}, it is not hard to show that
\begin{equation}
    F_Q[\ket{\alpha},\hat{n}^2]=4(\Delta \hat{n}^2)\rvert^2_\alpha=4(4\bar{n}^3+6\bar{n}^2+\bar{n}),
\end{equation}
and we observe a $\bar{n}^3$ scaling of the QFI. This results in {\em super-Heisenberg} scaling for the phase estimations where $\Delta\theta\sim 1/\bar{n}^{3/2}$. Experimentally, super-Heisenberg precision has only been demonstrated using a nonlinear atomic interferometer \cite{Sewell2014UltrasensitiveInterferometer}, and using many-body couplings in NMR \cite{Nie2018ExperimentalState}. Before describing how to achieve super-Heisenberg scaling for phase estimation of an {\em unknown} phase $\theta$, using the protocol shown in Fig.~\ref{fig:abstractcircuit1}, we first outline how the QFI generalises to the outputs of the circuit shown in Fig.~\ref{fig:abstractcircuit1}.  

\subsection{Quantum Fisher Information of a post-measurement ensemble}\label{QFIensemble}
In the above we considered the QFI associated with a parameter $\theta$, which modulates a {\em unitary} evolution of the initial fiducial state. In the scheme described in Fig.~\ref{fig:abstractcircuit1}, however, the final state of single run of the quantum circuit returns: $(m(\theta), \hat{\rho}_P(m,\theta))$, where $m$ the classical measurement result and $\hat{\rho}_P(m,\theta)$ is the corresponding conditional state of the Probe mode associated with that measurement result. As the measurement results $m$ are random from run-to-run, the resulting average quantum evolution that the Probe suffers is non-unitary and we have to generalise the Cramer-Rao bound and Quantum Fisher Information to this ensemble case.  To handle this  we apply the approach described in Ma \emph{et al.} \cite{Ma2019ImprovingTrajectories}, which we now briefly summarize.

Ma \emph{et al.} consider an extended system consisting of the system of interest (our Probe mode), and an environment to which the system couples unitarily (our Ancilla mode), and define the full density matrix of the extended system as
 $   \hat{\rho}_{\text{ext}}(\theta) = \hat{U}_{\text{ext}}(\theta) (|E_0 \rangle \langle E_0 | \otimes \rho_0) \hat{U}^{\dagger}_{\text{ext}} (\theta)
    \label{eqExtendedSystem},
$
where $|E_0\rangle$ is the $\theta$-independent initial state of the environment and $\hat{\rho}_0$ is the initial state of the system. The environment is then traced out in a $\theta$-independent basis $\{ |E_l\rangle \}$, and the reduced density matrix for the system alone can be written as
\begin{equation}
     \hat{\rho}_{\text{sys}}(\theta)= {\text{Tr}}_E\, \hat{\rho}_{\text{ext}}(\theta) = \sum_l \hat{\Pi}_l(\theta)\, \hat{\rho}_0\, \hat{\Pi}^{\dagger}_l(\theta) = \sum_l \hat{\tilde{\rho}}_l(\theta)\label{rhosys}
\end{equation}
where
$
    \Pi_l (\theta) = \langle E_l| U_{\text{ext}}(\theta) | E_0 \rangle
$,
are Kraus operators operating on the system. Essentially the $\hat{\tilde{\rho}}_l(\theta)$ are a set of quantum trajectories that occur with probability $P_l(\theta) = {\text{Tr}}[ \hat{\tilde{\rho}}_l(\theta)]$, and in our case they will be pure states. In this picture, an optimal measurement of the quantum system to estimate  $\theta$ yields the generalised Quantum Fisher Information
\begin{equation}
    {\mathbb{F}} = F[\{P_l\}] + \sum_l\, P_l\, F_Q[\hat{\bar{\rho}}_l],
    \label{eqFisherInfo1}
\end{equation}
where $F$ is the classical Fisher information of the distribution of measurement results given by 
$
    F[\{P_l\}] = \sum_l (\partial_{\theta} P_l)^2/P_l\;,
    \label{eqClassicalFisher1}
$ 
 $\hat{\bar{\rho}}_l = \hat{\tilde{\rho}}_l / P_l$, is the {\em normalized reduced density matrix of the system conditioned on the measurement result $l$}, and $F_Q$, is the single instance QFI given above in (\ref{ec:qfi_sld}). 
Using this form of the QFI, the Cramer Rao bound is given by
\begin{equation}
    \overline{\Delta \theta}^2 = \sum_{l} P_{l} \left( \Delta \theta  \right) ^2  \geq \frac{1}{ \nu\, \mathbb{F}}
\end{equation}

In our protocol, the conditioned state of the system, $\hat{\bar{\rho}}_l$, corresponds to the normalised final density matrix of the Probe mode $\hat{\rho}_P(m)$, exiting from Stage III of the protocol in Fig.~\ref{fig:abstractcircuit1}. We obtain this by scaling the conditioned state by $P(m)$, the probability of our measurement returning a value $m$, i.e. $\hat{\bar{\rho}}(m) = \hat{\tilde{\rho}}(m) / P(m)$, where $\hat{\tilde{\rho}}(m)$, is the final un-normalised post-measurement Probe state at the completion of stage III in Fig.~\ref{fig:abstractcircuit1}. We can now work out the generalised QFI to be
\begin{equation}
    {\mathbb{F}} = \int _{-\infty} ^{\infty}  \frac{(\partial_{\theta} P(m))^2} {P(m)} \, { dm} + \int _{-\infty} ^{\infty} P(m)\, {F_Q} [\hat{\bar{\rho}}(m)] \, { dm}
    \label{eq:TotalFisher}
\end{equation}
where the first term is the standard generalised classical Fisher information ${\mathbb F}_C$, and the second term is the generalised quantum Fisher information ${\mathbb F}_Q$.
To calculate ${F_Q} [\hat{\bar{\rho}}(m)]$ we make use of the fact that if we consider the {\em normalised} post-measurement Probe state,  it is in a pure state (see (\ref{ec:output2})), allowing us to use (\ref{ec:qfi_pure}). In the analysis below we find that the classical portion of ${\mathbb{F}}$ (which only depends on $P(m)$), is negligible when compared with the second term, the ensemble averaged QFI, and we will typically focus on the latter.
Finally, we mention that the last unitary in stage III in Fig.~\ref{fig:abstractcircuit1} is a correction unitary depending on the measurement result. This unitary introduces an additional $\hat{U}^C_l$ in (\ref{rhosys}) and we get $\hat{\rho}_{\text{sys}}(\theta)= {\text{Tr}}_E\, \hat{\rho}_{\text{ext}}(\theta) = \sum_l \hat{U}^C_l\hat{\Pi}_l(\theta)\, \hat{\rho}_0\, \hat{\Pi}^{\dagger}_l(\theta)\hat{U}^{C\,\dagger}_l = \sum_l \hat{\tilde{\rho}}_l(\theta)$. This does not alter $P(m)$, and except for the change $\hat{\bar{\rho}}_l\rightarrow \hat{\bar{\rho}}_l^C\equiv \hat{U}^C_l\hat{\bar{\rho}}_l\hat{U}^{C\,\dagger}_l$, the above derivation of the generalised QFI proceeds unchanged. We can thus evaluate the generalised QFI of the corrected ensemble using (\ref{eq:TotalFisher}), using  $\hat{\bar{\rho}}_l^C$. In what follows we  drop the $\hat{\tilde{\rho}}$ and $\hat{\bar{\rho}}$,  notations for post-measurement un-normalised or normalised states, referring instead to state vectors $\Ket{\Psi}$, which possess  non-unit or unit norms.

\vspace{1cm}

\subsection{Super-Heisenberg Metrology}
\label{superh}
In Section \ref{basicidea}, we observed that the conditioned post-measurement quantum state, $\ket{\psi_{III}^M}$, is similar to Kerr-type evolution of the initial probe state e.g. $|\psi\rangle_{kerr}\sim \exp(i\chi \hat{n}^2)|\alpha\rangle$. It is well known that quantum estimation of the strength of the deterministic Kerr can be performed with so-called {\em super-Heisenberg} scaling in the precision e.g. $\Delta \chi\sim 1/\bar{n}^{3/2}$ \cite{Luis2004NonlinearLimit}, and we now explore how this can be used to perform {\em super-Heisenberg} metrology of the rotation angle $\theta$, appearing in stage II of the protocol shown in Fig.~\ref{fig:abstractcircuit1}. In particular we assume we wish to estimate the value of an unknown small deviation, $\delta \theta$, of this angle from a pre-known bias value e.g. $\theta=\theta_0+\delta\theta$. The protocol involves a position measurement which returns a classical result $m$, with an associated probability distribution $P(m)$, and we will be interested in exploring how the generalised QFI (\ref{eq:TotalFisher}) scales with the resource $\bar{n}$, the expected photon number of the input Probe state. We recall from Section (\ref{quick}) that shot-noise scaling of $\theta$ corresponds to ${\mathbb{F}}\sim \bar{n}$, Heisenberg scaling corresponds to ${\mathbb{F}}\sim \bar{n}^2$, while {\em super-Heisenberg} corresponds to ${\mathbb{F}}\sim \bar{n}^\nu$ with $\nu>2$. Below we will find that the use of the {\em correction unitary}, $U_C$, shown at the end of stage III in Fig.~\ref{fig:abstractcircuit1}, plays a crucial role in achieving super-Heisenberg scaling. Without this correction our protocol achieves a precision worse than shot-noise. 




To begin, we note that the normalised state $|\Psi^M_{III}\rangle$ of the probe after measurement is pure, which enables us to calculate the Fisher information using (\ref{ec:qfi_pure}). Before doing that we first study the form of the post-measurement state. 

Taking the initial probe state to be $|\Psi_0\rangle_P$, which from (\ref{ec:output1}) and using the homomorphic transformation between the $(\beta,\sigma)\leftrightarrow (\theta, r)$ parameterizations yields 
\begin{equation}
|\Psi_{III}^{M}\rangle = N'_{IIIM}\exp \left[ - \frac{\sigma^2 (g \hat{n} - m)^2)} {2(1+2i\beta \sigma^2)} \right] |\Psi_0\rangle_P=N'_{IIIM}\exp\left[-f(m,\theta) \, (1-\frac{g}{m}\,\hat{n})^2\right]\,\ket{\Psi_0}_P,
\label{aftermeasurement}
\end{equation}
where 
\begin{equation}
    f(m,\theta)=\frac{m^2}{4}\,\frac{1-i\, e^{2r}\cot(\theta)}{e^{2r}-i\,\cot(\theta)}. \label{fullf}
\end{equation}
If we now consider the large squeezing limit $r\gg 1$, we find $f(m,\theta)\sim -i\, m^2/4\,\cot(\theta)$, and we curiously discover that the effect on the probe mode by the circuit is completely unitary,
\begin{equation}
    |\Psi_{III}^{M}\rangle=\exp\left[ \frac{i}{4}\cot(\theta)\,(m-g\hat{n})^2\right]\ket{\Psi_0}_P, \label{strongsqueezing}
\end{equation}
and thus, in this large squeezing limit, the decoherence channel presented to the Probe mode post-measurement is unital, mapping the identity to itself. In this limit we observe that the random measurement result $m$, causes a random phase rotation, which more generally depends on $(r, \theta, g, m)$.  
It is this random phase factor that will prevent us from obtaining optimal scaling of the generalised QFI, as essentially this is information we are throwing away after each measurement. In the large squeezing limit, if we are able to apply a unitary correction operation $U_C=\exp[i g\, m\, \cot(\theta)\, \hat{n}/2]$ to the probe state to cancel this phase term, the unknown random measurement effect will be removed. There is, in addition, an additional global phase which is proportional to $m^2$, but this global phase cannot influence the generalised QFI as it is not a physical observable in experiments on the post-measurement state. If we assume we are able to apply the unitary correction $U_C$, the normalized post-corrected pure state in the general case of finite squeezing can be written as 
\begin{equation}
|\Psi^C_{III}\rangle = N(m,r,\theta) \exp \left[ f(\theta,\hat{n}) \right] |\Psi_0\rangle_P
\end{equation}
where
\begin{equation}
f(\theta,\hat{n}) = \frac{i}{2}\, g \, m\, \hat{n} \cot( \theta)  - \frac{m^2}{4}\left[ \frac{1-i\, e^{2r}\cot({\theta})} {e^{2r}-i\, \cot(\theta)} \right]\left( 1-\frac{g}{m}\hat{n}\right)^2,
\label{eqfoftheta}
\end{equation}
 and where the normalisation $N(m,r, \theta )$, is taken to be real.


In order to apply this phase compensation we note that as part of our protocol we already have assumed that we have access to an oracle in Stage II that applies a number-dependent phase shift $R(\varphi)\equiv \exp[i \varphi \hat{n}]$ to the Ancilla mode. As mentioned above, we are primarily interested in the estimation of a small unknown $\delta\theta$ about a bias value $\theta_0$, i.e. $\varphi=\theta_0+\delta\theta$.  
Since
\begin{equation}
\cot(\theta_0 + \delta \theta) = \cot\theta_0 -(1+\cot^2 \theta_0) \delta \theta + \ldots
\label{eqCotThetaExpansion}
\end{equation}
by reusing this oracle, but now operating on the Probe mode, we observe that although we cannot generate the exact $\cot(\theta)$ unitary correction, the oracle, when used with known values of $\varphi$, and the unknown value $\theta=\theta_0+\delta\theta$,  is capable of applying a compensation to first order in $\delta \theta$. In this case the exponent function $f$ is given by 
\begin{equation}
f(\delta\theta,\hat{n}) = \frac{i}{2} g m \hat{n} (\cot \theta_0 -(1+\cot^2 \theta_0)\delta\theta)  - \frac{m^2}{4} \frac{1-i e^{2r}\cot(\theta_0 + \delta\theta)} {e^{2r}-i \cot(\theta_0 + \delta\theta)} \left( 1-\frac{g}{m}\hat{n}\right)^2,
\label{eqfofthetaCompensated}
\end{equation}
where the unknown small parameter to be estimated is now $\delta\theta$.

To compute the generalised QFI ${\mathbb F}$, we need to evaluate 
\begin{eqnarray}
\langle d \Psi_{III}^C|d \Psi_{III}^C\rangle &=&
\left| \frac{dN}{N} \right|^2 + \left( \frac{dN}{N} \right)^{\ast} \left\langle df \right\rangle + \left( \frac{dN}{N} \right) \left\langle df \right\rangle^{\ast} + \left\langle df^{\dagger}  df  \right\rangle \\
\left| \left\langle \Psi^C_{III} |d \Psi^C_{III} \right\rangle \right|^2 &=&
  \left| \frac{dN}{N} \right|^2 + \left( \frac{dN}{N} \right)^{\ast} \left\langle df \right\rangle + \left( \frac{dN}{N} \right) \left\langle df \right\rangle^{\ast} + \left| \left\langle  df  \right\rangle \right|^2
\end{eqnarray}
where $d \equiv d/d(\delta\theta)$, and $f$ is given as in (\ref{eqfofthetaCompensated}). From this the pure state QFI is given by
\begin{eqnarray}
F_Q(\ket{\Psi_{III}^C}) &=& 4 \left( \langle d \Psi_{III}^C|d \Psi_{III}^C\rangle  -
\left| \left\langle \Psi^C_{III} |d \Psi^C_{III} \right\rangle \right|^2\right), \\
&=&4\left(
\langle 
df^{\dagger}df\rangle 
- \left| \langle  df\rangle \right|^2 
\right), \\
&=&F_Q[\ket{\Psi_{III}^C},f] ,
\label{eqQFIBasic} 
\end{eqnarray}
where the expectation values are taken with the respect to the normalized state $|\Psi^C_{III} \rangle$, with $\delta\theta=0$.
This expression can be calculated analytically using (\ref{eqfofthetaCompensated}), but while straightforward, the resulting expression is lengthy and for that reason we do not explicitly include it here. This pure state $F_Q$ is a function of $(r, \theta_0, g, m)$, and to obtain the right hand term of the generalised QFI in (\ref{eq:TotalFisher}), the generalised Quantum Fisher Information, we must find a weighted average of $F_Q$ over all  possible measurement outcomes $m$, yielding
\begin{equation}
 {\mathbb F}_Q(r, \theta_0, g) = \int_{-\infty}^{\infty}\, P(m)\, F_Q(r, \theta_0,g, m)\, dm, \label{FQ}
\end{equation}
where $P(m)\equiv \langle \Psi_{III}^M|\Psi_{III}^M\rangle$.
Assuming, as we did in Section \ref{basicidea}, that the input Probe state is a coherent state, $\ket{\Psi_0}_P=|\alpha\rangle$, with expected photon number $\bar{n}_P$, we find that although the expressions (\ref{FQ}), and  $P(m)$,  can be found analytically, they cannot be evaluated analytically except in the simplest cases and we thus evaluate them numerically. We note that this can be difficult, as one must check convergence of these numerical expressions both with the Fock number truncation, and also with the precision in numerical accuracy as the integrand in (\ref{FQ}) can oscillate rapidly in cases and is often vanishing outside a compact domain in $m$.

We are now in a position to examine the scaling of the generalised Fisher information $\mathbb F$  (\ref{eq:TotalFisher}) to estimate $\delta\theta$ using our protocol. We assume a coherent Probe state input $|\Psi_0\rangle_P = |\alpha\rangle$, with expected Fock number $\bar{n}_P$, and assume that while we cannot carry out a full $\cot(\theta)$ phase compensation as in 
(\ref{eqfoftheta}), we can cancel the phase to first order, as in (\ref{eqfofthetaCompensated}). 

\begin{figure}
    \begin{centering}
  \includegraphics[width=13cm]{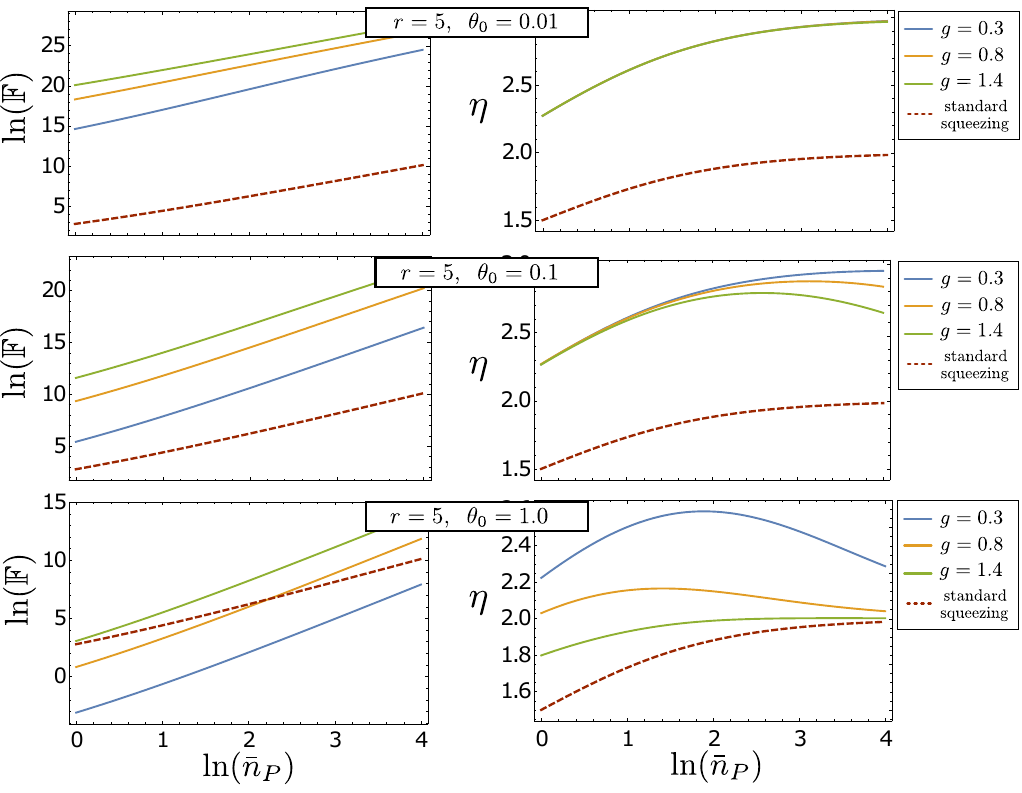}
        \end{centering}
        \caption{Plot of the generalised Fisher information for phase estimation ${\mathbb{F}}$ (left) and the exponent scaling $\eta$, of ${\mathbb{F}}\sim \bar{n}_P^\eta$, (right) as a function of the Probe mode mean photon number $\bar{n}_P$,  of our scheme, and for standard metrology using a squeezed input Probe state with an identical $\bar{n}_P$. We consider various cross-rotation coupling strengths $g$, Ancilla mode squeezing strengths $r$, and bias phase shifts $\theta_0$. The curves shown in the right panels are the slopes of the curves shown in the left panels. We take the horizontal axis to be the natural logarithm of $\bar{n}_P$, the mean photon number of the input Probe coherent state, and in the case of standard squeezing, it is the mean photon number of the input squeezed state. The curves colored {\bf (Blue, orange, green)}, correspond to {$\mathbf{ g=( 0.3,\,   0.8,\,  1.4)}$}, respectively, while the {\bf red curves} correspond to  {\bf standard squeezing}. Note that in the bottom right plot the blue, orange and green curves lie on top of each other. Super-Heisenberg scaling corresponds to $\eta> 2$, and from the right panels we observe many situations which achieve this for our protocol. In these numerics we set the Fock truncation to be $N_{trunc}=260$, and truncate the numerical integration along the $m-$axis to be within the domain $m\in[-900,900]$.}
        \label{fig:QFIscalingWithCompensationVariousTheta}
    \end{figure}
\begin{figure}
  \begin{centering}
  \includegraphics[width=14cm]{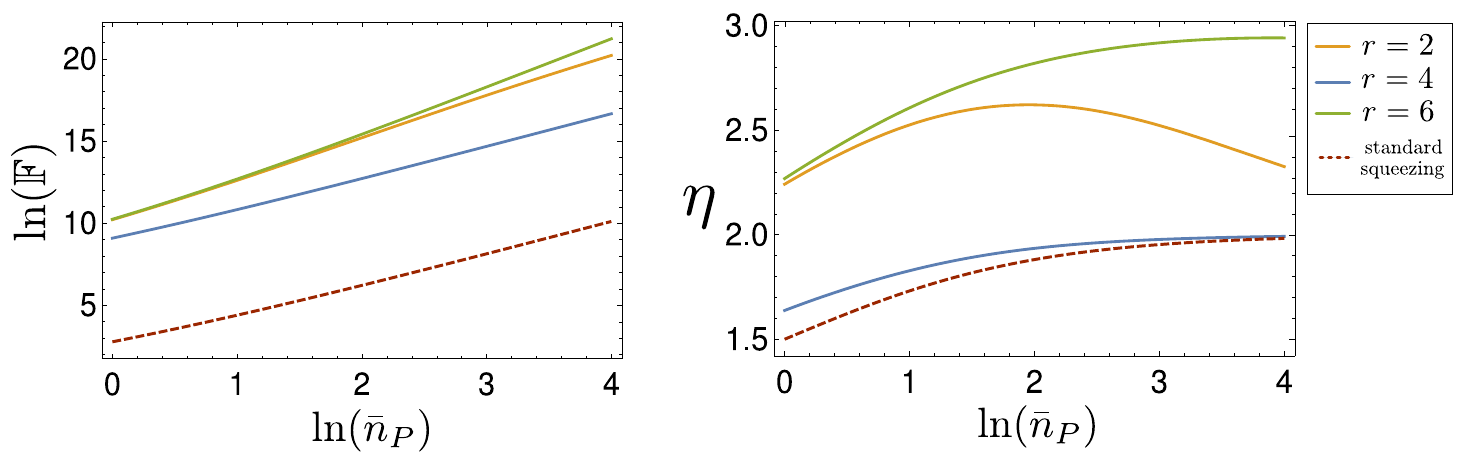}
  \end{centering}
  \caption{ Plot of the generalised QFI of the linearly corrected scheme for various values of Ancilla mode squeezing $r$,  compared to the case of Heisenberg scaling using a standard squeezed state with the same $\bar{n}_P$.  Curves colored {\bf ( orange, blue green)} correspond to $\mathbf{r=(2,\,4,\,6)}$, respectively, with $\theta_0 = 0.1, g=1.0$. The red curve corresponds to standard squeezing.}
  \label{fig:QFIscalingWithCompensationVariousR}
\end{figure}
\begin{figure}
        \begin{centering}
           \includegraphics[width=12cm]{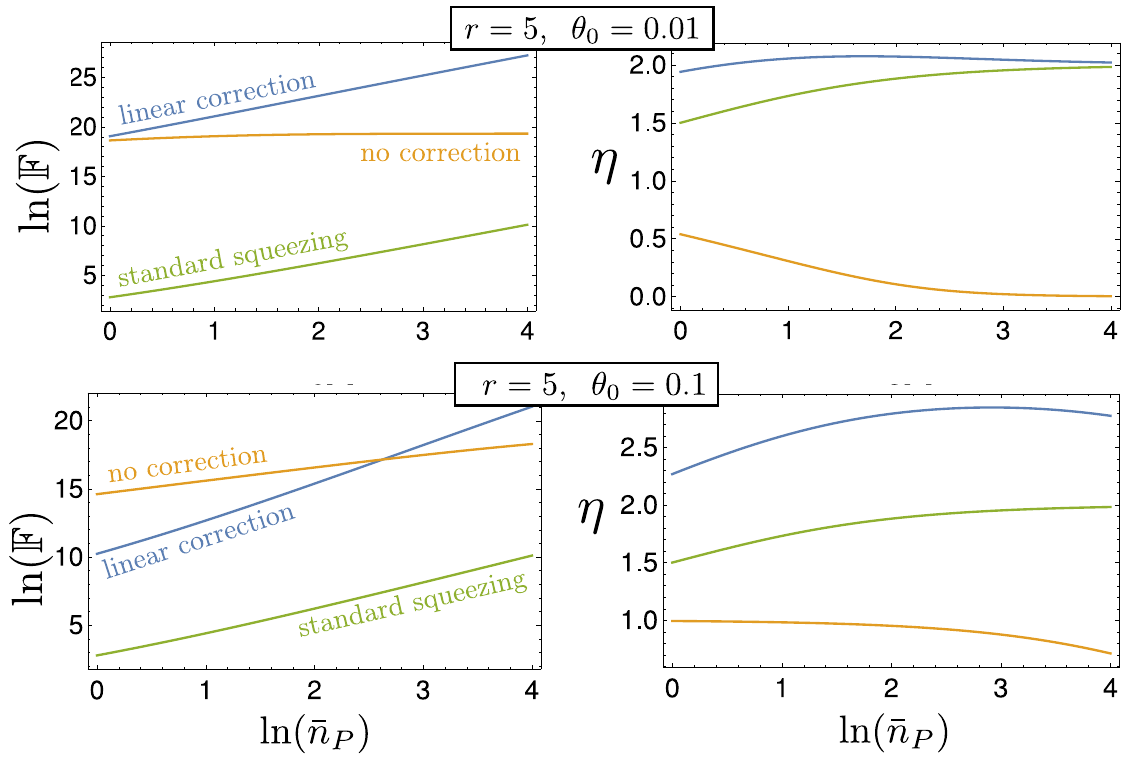}
        \end{centering}
        \caption{Comparison of how well the linear correction scheme does when compared to the non-corrected scheme. Shown is the generalised Fisher information  ${\mathbb{F}}$ (left column) and the scaling exponent $\eta$ (right column). The {\bf (orange, green, blue)} curves indicate {\bf (no correction, standard squeezing, linear correction)}, with $g=1.0$.}
        \label{fig:QFIscalingCompareCompensationAndNoCompensationTheta1e-0AndTheta1e-2}
    \end{figure}

We begin by examining how the quantum Fisher information ${\mathbb{F}}(r,\theta_0,g)$ scales with respect to  the average photon number $\bar{n}_P$ of the Probe. 
Figure~\ref{fig:QFIscalingWithCompensationVariousTheta} shows the behaviour of ${\mathbb{F}}$, for various values of $\theta_0$, coupling strength $g$, and squeezing $r$. We also graph the exponent $\eta$, given by ${\mathbb{F}} \sim \bar{n}_P^\eta$, by defining $\eta=d(\ln {\mathbb F})/d(\ln \bar{n}_P)$. Recall that the standard quantum limit Heisenberg and super-Heisenberg scaling corresponds to $\eta=1.0,\, 2.0,\, >2.0$, respectively.  During the numerical evaluations we discover that ${\mathbb F}_C\ll {\mathbb F}_Q$, and so ${\mathbb F}\sim{\mathbb F}_Q$ (\ref{eq:TotalFisher}).
From Fig.~\ref{fig:QFIscalingWithCompensationVariousTheta}, two things are clear: First, ${\mathbb{F}}$ is much higher than if we performed the estimation of $\delta\theta$ using the Ancilla mode alone when prepared in a standard squeezed state with the same value of $\bar{n}_P$. Second, the actual scaling with $\bar{n}_P$ is better than that found using a squeezed state resource, {\em and so the scaling in the Fisher Information is super-Heisenberg}. Using a squeezed state resource asymptotically approaches ${\mathbb{F}}\sim \bar{n}_P^2$ for large $\bar{n}_P$, whereas our scheme approaches ${\mathbb{F}}\sim \bar{n}_P^3$.
We also note the role of the bias angle $\theta_0$, as one may wish choose  values of the angle $\theta_0$ which maximises the Fisher information. 
Results for $\theta_0 = (0.01,\, 0.1,\, 1.0)\,{\rm rad}$, are shown in Figure~\ref{fig:QFIscalingWithCompensationVariousTheta}. From this we observe that as $\theta_0 \rightarrow 0$, we obtain a much higher absolute value of the generalised Fisher information ${\mathbb{F}}$, but the scaling of ${\mathbb{F}}$ with $\bar{n}_P$ weakens and does not hold over the same large range of $\bar{n}_P$.
We also study how the generalised QFI depends on the Ancilla mode squeezing $r$. This is shown in Figure~\ref{fig:QFIscalingWithCompensationVariousR}. As the squeezing $r$ is increased, we obtain both higher absolute Fisher information as well as better scaling, although the gains do saturate. 
One obvious question is how well the linear correction scheme works; that is, how well does the generalised QFI resulting from (\ref{eqfofthetaCompensated}) compare to the case when we just throw that information away and don't apply the correction unitary $U_C$ at all, i.e. taking the final state of the protocol to be (\ref{aftermeasurement})? This is plotted in Figure \ref{fig:QFIscalingCompareCompensationAndNoCompensationTheta1e-0AndTheta1e-2}. As before, a smaller $\theta_0$ results in a higher absolute ${\mathbb{F}}$ but worse scaling. We also see that without any  correction, in no case does the protocol scale better than the shot noise limit. However, with correction we achieve super-Heisenberg scaling and reach a total Fisher information greater than what can be obtained from using a standard squeezed state with the same $\bar{n}_P$.

In summary we have numerically evaluated the generalised Fisher Information for the linear corrected metrology protocol and find that for a range of coupling strengths $g$, and bias angles $\theta_0$, the protocol exhibits ${\mathbb F}\sim \bar{n}_P^\eta$, where $\eta>2$, or {\em super-Heisenberg} scaling. The central idea behind how this is achieved is essentially via the integral (\ref{doesthishaveaname}), which effectively squares $b$ in the exponential. By applying the cross-rotation operator $\exp(-ig\,\hat{p}_A\otimes \hat{n}_P)$ in Stage III, we ended up with a Kerr type evolution on the Probe mode $\sim \exp(ig^2/4 \cot(\theta)\,\hat{n}^2)$, in the limit of large Ancilla mode squeezing. Following this logic, if instead we applied the cross-mode operation $\exp(-ig\,\hat{p}\otimes \hat{A}_P)$, then we would end up applying the operation $\exp(ig^2/4\cot(\theta)\hat{A}^2)$ on the Probe mode for $r\gg 1$. We now briefly describe how this observation can be used to adapt the circuit shown in Fig. \ref{fig:abstractcircuit1}, to  {\em bootstrap up}, other metrology protocols to improve the scaling of their Fisher information, effectively surpassing their normal Heisenberg metrology limits.

\subsection{Bootstrap Protocol}\label{bootstrap}
As mentioned above, the protocol described in Fig.~\ref{fig:abstractcircuit1}, can essentially generate a Kraus operator on a target mode which is generated by the {\bf square} of the target operator in the two-mode {\em gate} $C_R$. In quantum metrology one effectively wishes to estimate a parameter $\theta$ which naturally appears in some unitary $U(\theta)=\exp(i\theta\, \hat{G})$, via the action of this unitary on a state $\ket{\psi_\theta}=U(\theta)\ket{\psi_0}$. The scaling of the QFI effectively is given by the uncertainty of $\hat{G}$ with respect to $\ket{\psi_0}$ (see Eq.~(\ref{ec:qfi_hamilt})). The protocol in Fig.~\ref{fig:abstractcircuit1} is effectively able to ``boot-up'' the power of $\hat{G}$ appearing in the unitary $U$, e.g. $\exp(i\theta\, \hat{G})\rightarrow \exp(i h(\theta)\, \hat{G}^2)$, where the function $h$ may be determined. The scaling of the QFI to estimate $\theta$ after this ``boot-up'' will  now be determined by the uncertainty of $\hat{G}^2$, which will scale with a larger power of the resource than $\hat{G}$. In the above we have used the Probe occupation number as the basis for our resource counting as the Probe mode is often the experimentally adjustable component, while the elements in the Ancilla are held fixed. Using this insight we now show how to adapt the protocol described in Section \ref{superh}, to be able to estimate the parameter generating displacements in phase space with an imprecision which scales better than the standard Heisenberg limit. Such displacement estimation is a crucial ingredient for many force sensing schemes. In this case $\ket{\psi_\kappa}\equiv \exp(i\kappa\,\hat{p})\ket{\psi_0}$, where the base state $\ket{\psi_0}$ has mean photon number $\bar{n}$. The Heisenberg limit for displacement measurements is achieved when $\ket{\psi_0}$ is a squeezed state or compass state \cite{Dalvit2006}, and in that case the QFI achieves a scaling of the form $F_Q(\kappa)\sim \bar{n}$.  By using the bootstrapped scheme we outline below, we argue we can achieve $F_Q(\kappa)\sim \bar{n}^3$, for the QFI in displacement sensing. This would yield an imprecision which scales as $\Delta\kappa\sim 1/\sqrt{\bar{n}^3}$, rather than $\Delta\kappa\sim 1/\sqrt{\bar{n}}$,  providing a vast improvement of the accuracy for displacement/force sensing. 

\begin{figure}
        \begin{centering}
           \includegraphics[width=15cm]{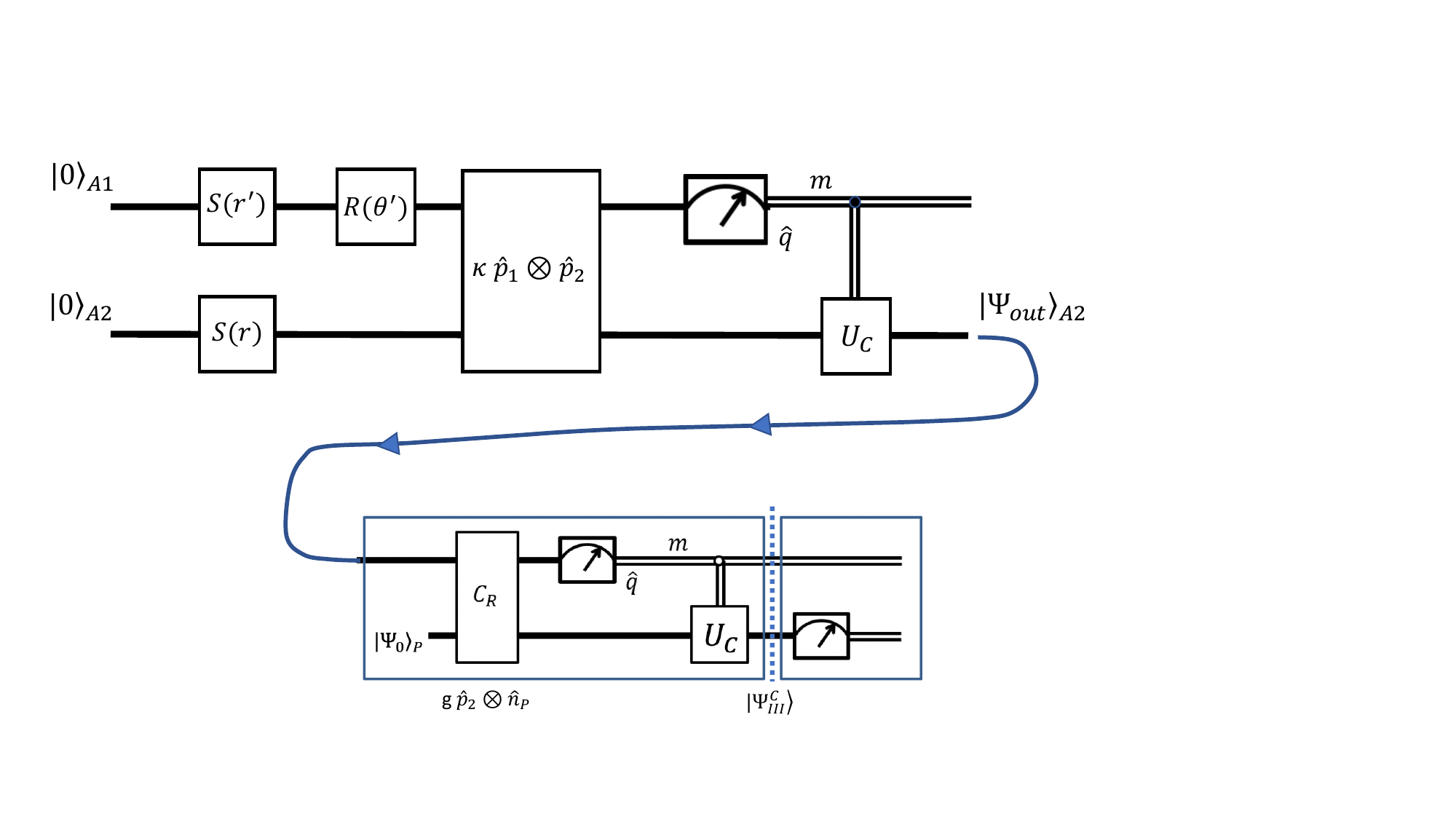}
        \end{centering}
        \caption{Schematic of the `Bootstrap' scheme to estimate the parameter $\kappa$, generating displacements in phase space.}
        \label{fig:Bootstrap1}
    \end{figure}

We focus on estimating the parameter $\kappa$, in the single mode displacement operator $D(\kappa)_{A1}=\exp(i\kappa \, \hat{p}_{A1})$. This operator displaces the mode $A1$ along the $\hat{q}_{A1}$ quadrature. We now assume we can access the two-mode gate $\exp(i\,\kappa\,\hat{p}_{A1}\otimes \hat{p}_{A2})$. Having access to this gate we can consider the ``Bootstrap'' circuit depicted in Fig.~\ref{fig:Bootstrap1}. The lower section of this circuit is Stage III and IV of the original circuit of Fig.~\ref{fig:abstractcircuit1}. The top circuit in Fig.~\ref{fig:Bootstrap1} is a replication of the original protocol but with some changes. Following the derivations in Section \ref{basicidea}, we see that in the large squeezing limit for mode A1, i.e. when $r^\prime \gg 1$, and when $\theta^\prime=\pi/4$, we have
\begin{equation}
    \ket{\Psi_{out}}_{A2}=\exp \left(\frac{i}{4}\,\kappa^2\,\hat{p}^2_2\right)\,S(r)\,\ket{0}_{A2}. \label{boot1}
\end{equation}
We now return to Eq.~\ref{IIIM}) and note that when $\ket{\Psi_{out}}_{A2}$ is input into Stage III and IV of the original circuit, Eq.~(\ref{IIIM}) remains unchanged except for $\beta\rightarrow -\kappa$. As the original protocol operating with linear correction is able to provide a generalised QFI for $\theta$ (or $\beta$) which scales as $\bar{n}^\nu$, with $\nu\in (2-3)$, we expect to find a similar scaling for the estimate of $\kappa$.

\section{Application to make Cat states and Compass states}\label{preparation}
We now return to the basic protocol outlined in Fig.~\ref{fig:abstractcircuit1}, but now assume we have full knowledge of all the parameters in the circuit, in particular the rotation angle $\theta$ in Stage II. In this situation, we study how well this circuit can be used to generate highly non-classical quantum states of the Probe mode.
Cat states and compass states \cite{Dalvit2006} have been proposed for quantum computation and for precision sensing of displacement \cite{Toscano2006Sub-PlanckMeasurements}. Such states exhibit fine detail in their Wigner functions and small displacements of such states become rapidly orthogonal. Here we show how our protocol allows deterministic preparation of high-fidelity cat and compass states.

To see this, we go back to the output state of the general protocol (\ref{ec:output2}), (\ref{eq:Ubeta})--(\ref{eq:Kcbetam}),
\begin{equation}
\Ket{\Psi_{\rm out}}=\mathcal{N}_0 \:U(\beta)\:U_c(\beta,m)\:K(\beta)\:K_c(\beta,m)\Ket{\Psi_0},
\label{ec:cats_psi_out}
\end{equation}

\noindent where $\mathcal{N}_0$ is a normalization factor and recall that the form of the conditioned unitary $U_c$ is 
\begin{equation}
U_c = \exp\left(i \frac{mg}{2}\frac{v}{u^2 + v^2}\hat{n}\right) \equiv e^{i \phi_c(m) \hat{n}},
\end{equation}
where we have made use of the holomorphic parameterisation (\ref{holomorpheqn}). As mentioned before, we will assume that $( \theta, r, g)$, are all controllable known parameters, and thus, although the precise values of the measurement result $m$ and the phase $\phi_c(m)$ are random in each execution of the protocol, these both are known precisely. As a consequence, the operation $U_c$ can be undone exactly independent of the initial state and the degree of squeezing in the Ancilla mode. Furthermore, in the high squeezing limit where $r\rightarrow\infty$ we have $K,\: K_c \sim \mathbb{I}$, and so the evolution ends up being dictated solely by the unconditioned unitary $U(\theta)$, 
\begin{equation}
U(\theta) = \exp\left(-i \gamma(\theta)\hat{n}^2\right),\ \mathrm{with}\ \gamma(\theta) = \frac{g^2v}{4(u^2 + v^2)}.
\label{ec:cats_gamma}
\end{equation}

Taking the initial Probe state to be a coherent state, we have that $\Ket{\Psi_{\rm out}}_P\simeq U(\theta) \Ket{\alpha}_P$ and thus a judicious choice of $\gamma(\theta)$ will yield states such as
\begin{eqnarray}
    \gamma(\theta)=\frac{\pi}{2} & \Rightarrow & e^{-i\frac{\pi}{2} \hat{n}^2}\Ket{\alpha}=\frac{e^{-i\frac{\pi}{4}}}{\sqrt{2}}\left(\Ket{\alpha}+i\Ket{-\alpha}\right)\equiv \Ket{\mathrm{cat}} \label{ec:cats_psicat} \\
    \gamma(\theta)=\frac{\pi}{4} & \Rightarrow & e^{-i\frac{\pi}{4} \hat{n}^2}\Ket{\alpha}=\frac{1}{2}\left[e^{i\frac{\pi}{4}}\left(\Ket{\alpha}-\Ket{-\alpha}\right)+\left(\Ket{i\alpha}+\Ket{-i\alpha}\right)  \right]\equiv \Ket{\rm compass}, \label{ec:cats_psicompass}
\end{eqnarray}
where we have dropped the Probe $P$ subscript. In the following we will study how well we can achieve these ideal target states in the case of large but finite squeezing in the Ancilla mode. 

\subsection{Choice of parameter $\theta$ and minimum squeezing requirements}

For finite squeezing, the nonunitary part of the evolution in Eq. (\ref{ec:cats_psi_out}) will impact how well we can prepare the desired target states. Before we analyze analyze such impact, we must determine under which circumstances it is \textit{a priori} possible to set the parameters of the evolution as in Eqs. (\ref{ec:cats_psicat})--(\ref{ec:cats_psicompass}). 
For this we can refer back to Eqs.~(\ref{holomorpheqn}) and (\ref{ec:cats_gamma}) to find
\begin{equation}
    \gamma(\theta) = \frac{g^2}{4}\frac{(1-e^{4r})\cot(\theta)}{e^{4r}+\cot^2(\theta)}
    \label{gammaeqn1}
\end{equation}
and for $r\gg 1$ we obtain $\gamma=-g^2\,\cot(\theta)/4$. For finite $r$ we notice that $\gamma(\theta)$ is an odd function about the value $\theta=\pi/2$, and 
obeys $\gamma(0)=\gamma(\pi/2)=0$ and thus it reaches a maximum absolute value $\gamma_{\rm max}=\gamma(\theta_c)$ in the interval $\theta\in [0,\pi/2]$ 
For fixed coupling strength $g$, the magnitude of $\gamma_{\rm max}$ increases with the amount of squeezing $r$ in the ancillary mode as depicted in Fig. \ref{fig:cats_gamma_zeta}.
This means that in order to be able to set the protocol parameters as required by Eqs. (\ref{ec:cats_psicat})-(\ref{ec:cats_psicompass}), a certain minimum squeezing value is required.  
One can derive an expression for $\gamma_{\rm max}$, which leads to the following squeezing thresholds (obtained numerically)
\begin{eqnarray}
\Ket{\rm compass}: &\   \gamma_{\rm max}/g^2 \geq \frac{\pi}{4} \Leftrightarrow r>0.93 \\ \Ket{\rm cat}: &\  \gamma_{\rm max}/g^2 \geq  \frac{\pi}{2}\Leftrightarrow r>1.27
\end{eqnarray}

In the remainder of this Section we take $g=1$ without loss of generality, and consider values of $r$ equal or greater than these threshold values, which allows us to set $\theta=\theta_*$ such that $\gamma(\theta_*)=\frac{\pi}{4}$ or $\frac{\pi}{2}$, depending on the target state.

\subsection{Effects of nonunitary backaction and conditioned state preparation fidelity}

\begin{figure}
    \centering
  \includegraphics[width=0.75\linewidth]{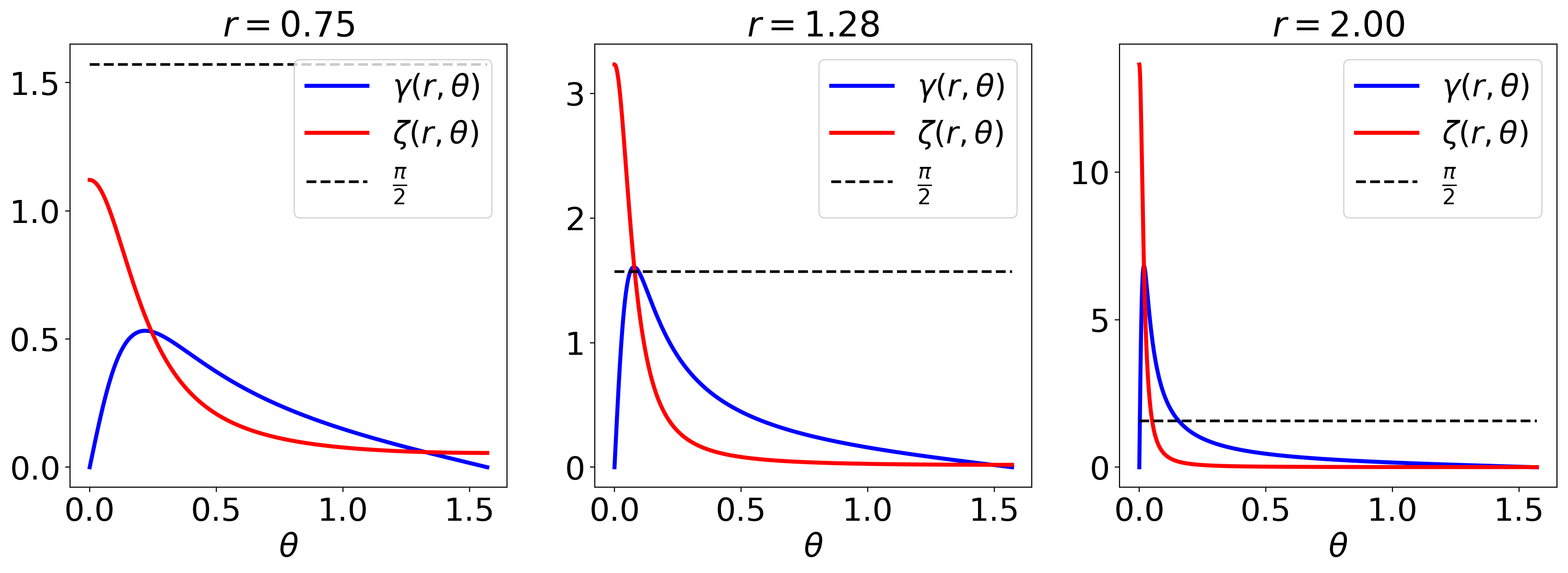}
\caption{ Plots of functions $\gamma(\theta)$ and $\zeta(\theta)$ defined in Eqs. (\ref{ec:cats_gamma}) and (\ref{ec:cats_zeta}) as a function of parameter $\theta$ and for different values of the squeezing parameter $r$ (here $g=1$). The parameter $\zeta(r,\theta)$ describes the non-unitary part of the process. We observe that $\gamma=\pi/2$ ($\pi/4$) can be achieved only if $r \gtrsim 1.27$ ($0.93$).  }
    \label{fig:cats_gamma_zeta}
\end{figure}

We now turn to analyzing how well our protocol is able to prepare to the target states of interest. We first define the (conditioned) state fidelity $\mathcal{F}_m$ as
\begin{equation}
    \mathcal{F}_m = \lvert \BraKet{\rm target}{\Psi_{\rm out}}\rvert^2
\end{equation}

\noindent where $\Ket{\rm target}=e^{-i \gamma(\theta_*) \hat{n}^2}\Ket{\alpha}$ refers to the target states in Eqs. (\ref{ec:cats_psicat}) and (\ref{ec:cats_psicompass}), and recall that $\theta_*$ is chosen to give $\gamma(\theta_*)=\frac{\pi}{2}$ or $\frac{\pi}{4}$ (depending on the particular target). Using Eq. (\ref{ec:cats_psi_out}) the fidelity can be cast directly in terms of the operators of interest,
\begin{equation}
    \mathcal{F}_m = \lvert \mathcal{N}\rvert^2 \lvert \Bra{\alpha}{U(\theta_*)^\dagger U(\theta_*)}K(\theta_*)K_c(m,\theta_*)\Ket{\alpha}\rvert^2 \\
    = \frac{\lvert \Bra{\alpha} K(\theta_*)K_c(m,\theta_*)\Ket{\alpha}\rvert^2}{\Bra{\alpha} K(\theta_*)^2 K_c(m,\theta_*)^2\Ket{\alpha}}
    \label{ec:cats_fidem1}
\end{equation}
\noindent where we have used that $\BraKet{\Psi_{\rm out}}{\Psi_{\rm out}}=1$. To analyze the effects of finite squeezing we recast the nonunitary part of Eq. (\ref{ec:cats_psi_out}) in the following form
\begin{equation}
K(\theta)K_c(m,\theta) = \exp\left[- \zeta(\theta) \hat{N}_m\right]\mathrm{with}\ \hat{N}_m\equiv \hat{n}^2  - \frac{2m}{g} \hat{n}\ \mathrm{and}\  \zeta(\theta)=\frac{g^2 u}{4(u^2+v^2)},
\label{ec:cats_zeta}
\end{equation}

\noindent where the function $\zeta(\theta)$, complementary to $\gamma(\theta)$, monotonically decays from $\zeta(0)=g^2e^{2r}/4$ to $\zeta(\frac{\pi}{2})=g^2e^{-2r}/4$ and is shown in Fig. \ref{fig:cats_gamma_zeta} for various values of $r$. For a given level of squeezing, setting $\theta_*$ as described above fixes the value of $\zeta(\theta_*)$, which for large $r$ we expect to behave like $\zeta(\theta_*)\equiv \zeta \sim e^{-2r} \ll 1$. We can then expand the operator in Eq. (\ref{ec:cats_zeta}) in powers of $\zeta$, $\exp(- \zeta \hat{N}_m)\simeq 1 - \zeta \hat{N}_m + \zeta^2\hat{N}_m^2/2$. Replacing this in Eq. (\ref{ec:cats_fidem1}) we obtain
\begin{equation}
    \mathcal{F}_m \simeq \frac{\left( 1 - \zeta \langle \hat{N}_m\rangle + \frac{\zeta^2}{2} \langle \hat{N}_m^2\rangle\right)^2 }{1 - 2\zeta \langle \hat{N}_m\rangle + 2\zeta^2 \langle \hat{N}_m^2\rangle},
\end{equation}
\noindent where the expectation values are taken over the initial coherent state $\Ket{\alpha}$. Keeping the leading order contribution we finally get
\begin{equation}
    \mathcal{F}_m \simeq 1 - \zeta^2\left(\langle \hat{N}_m^2\rangle - \langle \hat{N}_m\rangle^2\right) = 1 - \zeta^2 (\Delta \hat{N}_m)\rvert^2_{\alpha}.
    \label{ec:cats_fidem2}
\end{equation}

\noindent 

\subsection{Average fidelity and numerical results}

The fidelity in Eq. (\ref{ec:cats_fidem2}) still depends on the random measurement outcome $m$, which in turn is a function of the parameters of the protocol, including the squeezing $r$ in the Ancilla mode. In order to obtain a clearer picture about the target state preparation fidelity in our protocol, we consider the average (unconditioned) fidelity
\begin{equation}
    \mathcal{F}_{\rm avg} = \int dm\: P(m)\: \mathcal{F}_m
    \label{ec:cats_fideavg1}
\end{equation}
\noindent where $P(m)$ is the probability distribution associated with the random measurement outcome $m$, i.e.
\begin{equation}
    P(m) = \mathcal{A}{\Bra{\alpha} K_0(\theta_*)^2 K_c(m,\theta_*)^2\Ket{\alpha}}=\sqrt{\frac{2\zeta}{\pi g^2}} e^{-|\alpha|^2}\sum\limits_n \frac{|\alpha|^{2n}}{n!}\exp\left[-2\zeta\left(n-\frac{m}{g}\right)^2\right].
    \label{ec:cats_pm1}
\end{equation}

Since the conditioned fidelity is quadratic in $m$, in order to compute the average fidelity we need to compute the first two moments of the this distribution, $\overline{m}$ and $\overline{m^2}$. These can be computed easily thanks the Gaussian form of each term in Eq. (\ref{ec:cats_pm1}). For the mean value, we have 
\begin{eqnarray}
    \overline{m}=\int m\:P(m)\:dm &=& \sqrt{\frac{2\zeta}{\pi g^2}} e^{-|\alpha|^2}\sum\limits_n \frac{|\alpha|^{2n}}{n!} \int m\: \exp\left[-2\zeta\left(n-\frac{m}{g}\right)^2\right]dm \\
    &=& g e^{-|\alpha|^2}\sum\limits_n n \frac{|\alpha|^{2n}}{n!} = g e^{-|\alpha|^2}\sum\limits_m \frac{|\alpha|^{2m}}{m!}|\alpha|^2 = g|\alpha|^2.
\end{eqnarray}

Using similar techniques we can calculate the second moment, for which we obtain
\begin{equation}
\overline{m^2} = g^2\left(\frac{1}{4\zeta} +|\alpha|^2(1+|\alpha|^2)\right).
\end{equation}

With these results we can combine Eq. (\ref{ec:cats_fidem1}) with Eq. (\ref{ec:cats_fideavg1}) to obtain
\begin{equation}
    \mathcal{F}_{\rm avg} = 1 - \zeta|\alpha|^2 + \mathcal{O}(\zeta^2).
    \label{ec:cats_fideavg2}
\end{equation}

Notice that, due to the $\zeta^{-1}$ dependence of $\overline{m^2}$, the leading order contribution of the average fidelity is actually $\mathcal{O}(\zeta)$. The asymptotic expression in Eq. (\ref{ec:cats_fideavg2}), valid for large squeezing $r$, is the main result of this Section. It shows that, in this regime, we expect the fidelity of cat and compass state preparation to increase exponentially with the Ancilla squeezing $r$, since $\zeta \sim e^{-2r}$. For fixed squeezing, however, the average fidelity drops linearly with the mean photon number of the original Probe coherent state $\overline{n}=|\alpha|^2$. 

\begin{figure}
    \centering
  \includegraphics[width=0.9\linewidth]{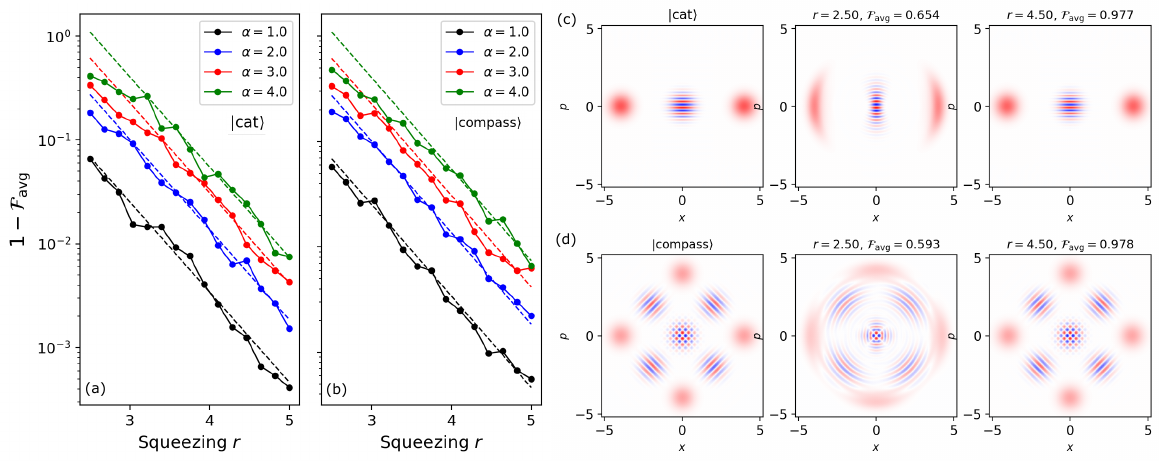}
\caption{ Cat and compass state generation with the proposed protocol. (a) - (b) show the average state infidelity $1-\mathcal{F}_{\rm avg}$ computed numerically as a function of the squeezing magnitude $r$ of the ancillary mode. Each value is obtained by averaging over 50 runs of the protocol. Results are shown for different values of initial coherent state amplitude $\alpha$, and for the cases where the target state is (a) $\lvert \rm{cat}\rangle$ and (b) $\lvert \rm{compass}\rangle$. Dashed lines indicate the leading order analytical estimate of Eq. (\ref{ec:cats_fideavg2}). (c) and (d) Wigner function of the ideal target states (left plots), and of the states generated by the protocol when $r=2.5$ (center) and $r=4.5$ (right). Results for Wigner functions are obtained from averages over 10 realizations of the protocol in each case, and the value of the coherent state amplitude is set to $\alpha=4$.  }
    \label{fig:cats_numerics}
\end{figure}

In order to test these results, we performed numerical simulations of the cat and compass state preparation protocol. Results are shown in Fig. \ref{fig:cats_numerics}. In (a) and (b) we show the average infidelity $1-\mathcal{F}_{\rm avg}$ computed over 50 runs of the protocol, for different values of initial coherent state amplitude. The results clearly show that the infidelity drops exponentially with the squeezing magnitude $r$, with excellent agreement with the analytical result of Eq. (\ref{ec:cats_fideavg2}) even for moderate values of $r$. We actually observe that for small $r$ (large $\zeta$) fidelities are higher than those predicted by the leading order calculation, indicating that the protocol behaves better than expected in this regime. The actual states achieved by the protocol are depicted in Fig. (\ref{fig:cats_numerics}) (c)-(d), where we plot the Wigner functions of the resulting states for the ideal ($r\rightarrow\infty)$ and the finite squeezing cases.

\section{Physical implementation of our scheme} 
\label{sect:atomic_prot}
In the above we have described how to effectively generate non-linear dynamics via the protocol described in Fig.~\ref{fig:abstractcircuit1}. We showed how it can be used to perform precision measurement and to synthesise non-classical quantum states. We now outline a potential physical implementation of our protocol using atomic ensembles interacting with light through the Faraday effect. We describe the single atomic ensemble as a bosonic mode via the Holstein-Primakoff approximation~\cite{Holstein1940FieldFerromagnet}, making the assumption that the ensemble is made up of very many ($N\gg 1$) atomic spins, and that the state stored in the ensemble has a close to maximal $R^{\dagger} J_Z R$ expectation value, for some $R\in$ SU(2), and where $J_K=\sum_{\alpha=1}^N \sigma_K^{(\alpha)}$ is the component of the collective angular momentum in the direction $K$ ($K=X,Y,Z$). 
Spin basis states are mapped to bosonic fock states via
\begin{align}
    R^{\dagger} \ket{j,j-m}_Z \mapsto \frac{1}{\sqrt{m!}} \left(a^{\dagger}\right)^{m}\ket{0},
\end{align}
and spin observables to bosonic observables via
\begin{align}
    R^{\dagger} J_+ R
    &= 
    \sqrt{2j} \sqrt{1-\frac{\hat{a}^{\dagger} \hat{a}}{2j}} \hat{a} 
    \approx
    \sqrt{2 j} \hat{a},
    \\
    R^{\dagger}J_- R 
    &= 
    \sqrt{2j} \hat{a}^{\dagger} \sqrt{1-\frac{\hat{a}^{\dagger} \hat{a}}{2j}} 
    \approx
    \sqrt{2 j} \hat{a}^{\dagger},
    \\
    R^{\dagger} J_Z R
    &=
    j-\hat{a}^{\dagger} \hat{a},
\end{align}
where the approximation sign holds for large $j=N/2$.
Our protocol requires two fundamental operations. The first involves implementing an SU(2) rotation using a magnetic field via a Zeeman interaction~\cite{Deutsch2010QuantumSpectroscopy}, i.e. $\vec{B}\cdot\vec{J}$.  The second uses the Faraday interaction $H_F=\chi J_Z S_Y$,
to couple the $J_Z$ component of the ensemble's angular momentum to polarization degree of freedom of light, which is described in terms of the Stokes vector operators: $S_{j} = \frac{1}{2} \mathbf{a}^{\text{H}} \sigma_{j} \mathbf{a}$ (here we use $\mathbf{a} = (\hat{a}_{H}, \hat{a}_{V})^{\text{T}}$ and $\mathbf{a}^{\text{H}} = (\hat{a}^{\dagger}_{H}, \hat{a}^{\dagger}_{V})$).

\begin{figure}
    \includegraphics[width=0.4\linewidth]{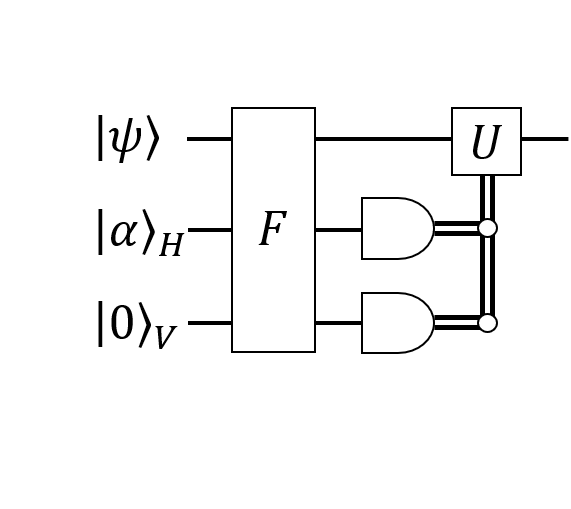}
    \caption{Gadget for measuring the spin-$Z$ direction of the ensemble in state $\ket{\psi}$. This is coupled to a pair of light modes (one for each polarization), by the Faraday interaction $H_F=\chi J_Z S_Y$. Then we use photodetectors to measure how much the light's polarization has shifted, thereby approximating the $Y$-component of angular momentum of $\ket{\phi}$. }
    \label{fig:meassq}
\end{figure}
\begin{figure}
    \centering
    \includegraphics[width=0.5\linewidth]{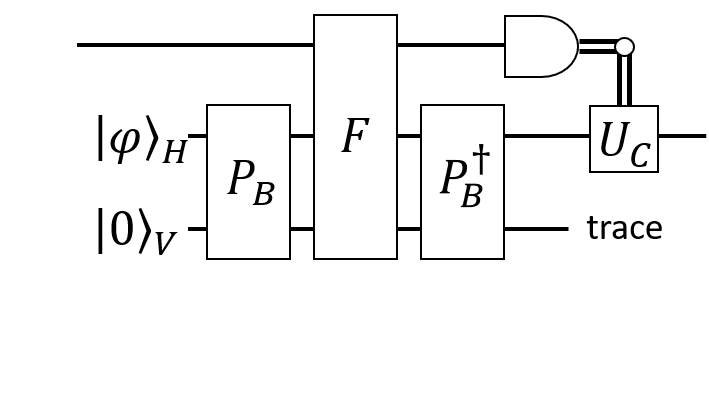}
    \caption{ We teleport the information from the ensemble onto the optical state in the $H$ mode. We can put this mode in any input state $\ket{\varphi}$, (which might be in a coherent state), and we do a $C_R$ operation by sandwiching the Faraday interaction between a pair of beamsplitters and rotations. Finally, we measure the ensemble along the $Z$ direction, which gives us information about the $\hat{q}$ quadrature in the Holstein-Primakoff approximation. Note that this measurement may itself require coupling to additional light modes. We also allow for a post-measurement unitary conditioned on the outcomes. }
    \label{fig:stage3circuit}
\end{figure}

We can prepare the ensemble in a momentum squeezed state (in the Holstein-Primakoff approximation centered at the positive X axis, achieved by setting $R=e^{-i\pi J_Y /4}e^{-i\pi J_X /4}$), by implementing $H_F$ and subsequently measuring the change in the plane of polarization of the light, as shown in the circuit of Fig.~\ref{fig:meassq}. This produces a sequence of weak measurements of $J_Z$, which are represented by the following Kraus operator applied to the state~\cite{Jacobs2014QuantumApplications,munoz2020}:
\begin{align}
    K_m = \frac{1}{(2\pi\sigma^{2})^{1/4}} e^{-\frac{1}{4\sigma^{2}}(J_Z - m)^2},
    \label{ec:kraus_atom}
\end{align}
where $\sigma$ is the measurement resolution, which is related to the interaction strength $\chi$ and the measurement time $\Delta t$.
In the Holstein-Primakoff picture, this will correspond to a displaced momentum squeezed state. We transform this into a position-squeezed vacuum state by correcting the measurement-dependent displacement using
$\exp\left[i t \frac{ J_Y }{2\sqrt{j}} \right]
    \mapsto \exp[-i t \hat{p}]$, followed by a $90^\circ$ rotation about the X axis, which implements a Fourier transform in the Holstein-Primakoff mode picture.
We can squeeze the input substantially if $e^{-2r} \approx \sigma^{2}/j \ll 1$.
After the preparation of a squeezed state, the ensemble interacts with the signal present in the magnetic field via the Zeeman interaction
\begin{align}
    B J_X = B R^{\dagger}J_Z R \mapsto B(j-\hat{a}^{\dagger}\hat{a}).
\end{align}
 In the Holstein-Primakoff picture of the ensemble as a bosonic mode, this acts as a single-mode phase gate with $\theta=B$. We next use the Faraday interaction $H_F=\chi J_Z S_Y$, to implement the $C_R$ gate (which is given in Fig.~\ref{fig:abstractcircuit1}, by $C_R=\exp(i\,g \hat{p}_{A}\otimes\hat{n}_P$). We do this as shown in Fig.~\ref{fig:stage3circuit}.
We use a polarizing beamsplitter $P_B=\exp[i \pi (\hat{a}_{H}^{\dagger}a_{V} + \hat{a}_{V}^{\dagger} a_{H})/2]$ to rotate the Stokes bases such  that $P_B S_{Y} P_B^{\dagger} = S_{Z} = \hat{a}^{\dagger}_{H}\hat{a}_{H} - \hat{a}^{\dagger}_{V} \hat{a}_{V}$. The final step is to ensure mode $V$ is in the vacuum state, so that we can ignore $\hat{a}^{\dagger}_{V}\hat{a}_{V}$. The final step in the atomic protocol is to perform the measurement of $\hat{q}$, which is achieved by $J_Y$. 

\section{Conclusion}
It is well known that generating large optical nonlinearities is exceedingly difficult and so far has primarily been achieved using superconducting quantum optical circuits at microwave frequencies. Finding methods to generate nonlinearities at optical frequencies opens up a wide range of paths for research and applications. In this work we presented a new protocol that is able to generate an optical nonlinearity via a conditional linear operation and measurement with feedback. Using this nonlinearity we show how to deterministically generate highly nonlinear quantum states of the Probe mode which can be very pure. We also show how to engineer the degree of the nonlinearity to depend on unknown parameters e.g. rotation angle $\theta$. Estimations of this parameter via this nonlinearity can be made with an imprecision which can scale as $\sim 1/\bar{n}^{3/2}$, beating the Heisenberg limit. Perhaps even more useful is the so-called bootstrap method, where we show how to engineer this optical nonlinearity to depend on a wider variety of parameters, e.g. the parameter associated with an unknown displacement, whose estimation is a central task in force metrology.  We finally describe a physical setup, using the Faraday interaction of light with atomic ensembles, to implement our protocol. In this case the unknown rotation angle could be generated by a physical magnetic field, thus allowing magnetometry with super-Heisenberg scaling in the sensitivity. At the heart of our protocol is the novel engineering of near-unitary Kraus operations whose random components can almost be completely compensated for. This opens up a completely new ability to perform near-deterministic Schrodinger evolution driven by measurements alone to achieve highly nonlinear and sophisticated quantum dynamics. 

\section*{Acknowledgements}
JT acknowledges funding from the Australian Research Council Centre of Excellence in Engineered Quantum Systems CE170100009. MR acknowledges support from Grant UNAM-DGAPA-PAPIIT IG101421. This  work  was partially  supported  by  the  U.S. National  Science  Foundation Grant  No. 1630114.

\bibliography{jasonsnonlinearmetrologymendeleygroup}












\end{document}